\documentclass{amsproc}

\theoremstyle{definition}

\theoremstyle{remark}

\numberwithin{equation}{section}

\begin{document}

\def\ben{\begin{equation}}
\def\een{\end{equation}}
\def\bea{\begin{eqnarray}}
\def\eea{\end{eqnarray}}

\def\cO{{\mathcal{O}}}
\def\cT{{\mathcal{T}}}
\def\tcO{\tilde{\mathcal{O}}}
\def\bA{{\bf A}}
\def\bC{{\bf C}}
\def\bX{{\bf X}}
\def\bY{{\bf Y}}
\def\bx{{\bf x}}
\def\bp{{\bf p}}
\def\bphi{{\bf \phi}}
\def\bB{{\bf B}}
\def\bF{{\bf F}}
\def\bI{{\bf I}}
\def\bO{{\bf O}}
\def\tpsi{{\tilde{\psi}}}
\def\sp{\star^\prime}
\def\tS{{\tilde{S}}}
\def\bag{{\bar g}}
\def\baG{{\bar G}}

\title{BULK COUPLINGS TO NONCOMMUTATIVE BRANES}

\author{Sumit R. Das}
\address{Tata Institute of Fundamental Research,
Homi Bhabha Road, Mumbai 400 005.}
\email{das@theory.tifr.res.in}

%    General info
%\subjclass{Primary 54C40, 14E20; Secondary 46E25, 20C20}
%\date{January 1, 1994 and, in revised form, June 22, 1994.}

\dedicatory{Based on talk at ``Strings 2001'', Mumbai, January 2001.}

%\keywords{Differential geometry, algebraic geometry}

\begin{abstract}
We propose a way to identify the gauge
invariant operator in noncommutative gauge theory on a D-brane with
nonzero B field which couples to a specific supergravity mode in the
bulk.  This uses the description of noncommutative gauge theories in
terms of ordinary $U(\infty)$ gauge theories in lower dimensions. The
proposal is verified in the DBI approximation. Other authors have
shown that the proposal is also consistent with explicit string
worldsheet calculations. We comment on implications to holography.
\end{abstract}

\maketitle

\section{Introduction and Overview}
The study of coupling of bulk modes with usual D-branes have played a
crucial role in our understanding of black holes, holography and many
other features of string theory. Recent developments indicate that
D-branes with constant NSNS $B$-fields probe a very fruitful corner of
string theory. The low energy theory of such D-branes is
noncommutative gauge theory
(NCGT)\cite{connes,ncym,seibergwitten}. 
In this talk I will discuss how to
couple such D-brane to bulk fields and its implications to holography,
based on work with S.J. Rey and S.P. Trivedi.

Bulk modes must, of course, couple to gauge invariant
operators. However in NCGT, gauge transformations and translations are
intertwined. For example, consider an adjoint field $\phi
(x)$. Under a gauge transformation with
gauge parameter $\Lambda(x) = b_ix^i$ one has
\begin{equation}
\phi (x) \rightarrow e^{ib_ix^i}\star \phi (x) \star e^{-ib_ix^i} = \phi
(x^i+\theta^{ij}b_j)
\label{eq:aone}
\end{equation}
This shows that such a gauge transformation is equivalent to a
translation by the parameter $\theta^{ij}b_j$ and implies that there
are no gauge invariant operators which are local in position space.

Since the theory has translation invariance,
gauge invariant operators with definite momentum must exist. 
Such operators were
constructed by Iso, Ishibashi, Kawai and Kitazawa (IIKK)
\cite{iikk}. These are fourier transforms of {\em open Wilson lines},
with the separation between the base points being proportional to the
momentum.  In fact, noncommutative gauge theories have supergravity
duals \cite{hi,mr} -
this implies that there must be operators of definite momentum
which are dual to the supergravity modes \cite{dghosh,ghi}. 
In \cite{dasrey} it was
proposed that these operators are in fact the operators constructed by
IIKK and the fact that their size increases with momentum is reflected
in the dual supergravity. In \cite{ghi} it was proposed that the
relevant operators are in fact Wilson lines which are straight, with
operator insertions at the end point and in \cite{dw} other
classes of open Wilson lines with operator insertions along it were
considered. Various properties of these operators have been studied in
\cite{ambjorn}. Correlation functions of Wilson tails have an
interesting {\em universal} high energy behavior \cite{ghi,dkit1,rozali}.
It was, however, unclear how to obtain the precise form of
such Wilson lines with operator insertions which couples to a {\em
specific} supergravity mode.

In \cite{dastrivedi} we proposed that the way to do this is to use the
construction of NCGT from ordinary $U(\infty)$ gauge theories in lower
dimensions or matrix models \cite{mmodels,mmodelso} which was used to write
down these operators in the first place. The latter is the
theory of a large number of lower dimensional branes with no $B$
field. 
Suppose we know the linearized couplings of a set
of ordinary $Dp$ branes to supergravity backgrounds. Then the proposal
is to use the
above construction to find the couplings of these backgrounds to
noncommutative $D(p+2n)$ branes with noncommutativity in $2n$ of the
directions. We find that the resulting operators are straight Wilson
lines, which we call Wilson tails, with local operators {\em smeared
along them}. By the nature of this construction we obtain the operator
in the $\Phi = -B$ description \cite{seibergwitten}.
Such Wilson tails with smeared operators were constructed
in \cite{liu} and argued to be the correct operators from a rather
different point of view. Generalized star products \cite{star}
which have appeared in various contexts in NCGT make an appearance
here as well.

The form of the coupling of supergravity modes to a large number of
ordinary $Dp$ branes is not known in general - it is known in the low
energy limit \cite{taylor} or in the DBI approximation
\cite{dbi,taylor}. The couplings are consistent
with the symmetrized trace prescription of \cite{tseytlin}. To test
our proposal we work in the DBI approximation and construct the
couplings to noncommutative branes.  For a single brane in the DBI
approximation, one can also write down the coupling in terms of
ordinary gauge fields and the $B$ field on the brane by standard
methods.  For dilaton couplings, we show, to second order in the
noncommutative gauge potential but to all orders in the
noncommutativity parameter, that the coupling we propose is in fact
identical to that in terms of ordinary gauge fields and transformed by
the Seiberg-Witten map. It has been subsequently shown that the
proposal is also consistent with worldsheet calculations of couplings
to other NSNS fields in the low
energy limit \cite{worldsheet}, and to couplings of RR fields
\cite{rrcoupling}.

\section{NCGT from large-N Yang-Mills}

The fact that space-time can be encoded in $N = \infty$ matrix models
living at a single point has been known for some time : this is the
basis for Eguchi-Kawai models \cite{ek}. In the twisted Eguchi-Kawai
model \cite{tek} the space-time which emerges is in fact
noncommutative. However since planar diagrams of commutative and
noncommutative gauge theories are identical, the twisted model indeed
describes Yang-Mills theory at large $N$. In the present context, we
want to describe finite $N$ noncommutative gauge theories in terms of
$N = \infty$ matrix models, or $N = \infty$ gauge theories in lower
dimensions. The fact that the twisted Eguchi-Kawai model has
noncommutative space built in it comes as a bonus \cite{mmodels}. In
fact this is how branes arise in matrix models \cite{bmatrix}, both in
the BFSS \cite{bfss} and IKKT \cite{ikkt} versions.

Consider a $U(\infty)$ ordinary gauge theory in $(p+1)$ dimensions
with the usual gauge fields $\bA_\mu(\xi)~~,\mu=1,\cdots p+1$ and
$(9-p)$ scalar fields $\bX^I(\xi),~I=1,\cdots (9-p)$ in the adjoint
representation, together with their fermionic partners. In this paper
we will restrict ourselves to bosonic components of operators.
Consequently, fermions will not enter the subsequent discussion.  The
bosonic part of the action is
\begin{equation}
S = {\rm Tr}\int d^{p+1}\xi [\bF_{\mu\nu}\bF^{\mu\nu} + 
D_\mu \bX^I D^\mu \bX^J g_{IJ}
+ [\bX^I,\bX^J][\bX^K,\bX^L]g_{IK}g_{JL}]
\label{eq:six}
\end{equation}
where $g_{IJ}$ are constants and the other notations are standard. Boldface
has been used to denote $\infty \times \infty$
matrices. 

The action has a nontrivial classical solution
\begin{equation}
\bX^i (\xi)  =  \bx^i~~(i=1,\cdots 2n);~~~~~~
\bX^a  =  0~~(a=2n+1 \cdots 9-p);~~~~~~~\bA_\mu  =  0
\label{eq:seven}
\end{equation}
where the constant (in $\xi$) matrices $\bx^i$ satisfy
\begin{equation}
[\bx^i,\bx^j] = i\theta^{ij} \bI
\label{eq:eight}
\end{equation}
The antisymmetric matrix $\theta^{ij}$ has rank $p$ and $\bI$
stands for the unit $\infty \times \infty$ matrix. The inverse of the
matrix $\theta^{ij}$ will be denoted by $B_{ij}$

The idea is then to expand the various fields as follows.
\begin{equation}
\bC_i  =  B_{ij}\bX^j = \bp_i + \bA_i;~~~
\bX^a  =  \bphi^a;~~~
\bA_\mu  =  \bA_\mu
\label{eq:nine}
\end{equation}
where ${\bf p}_i  =  B_{ij}\bx^j$. 
We will expand any matrix ${\bf O}(\xi)$ as follows
\begin{equation}
{\bf O}(\xi) = \int d^{2n}k~{\rm exp}[i\theta^{ij}k_i{\bf p}_j]~
O(k,\xi)
\label{eq:twelve}
\end{equation}
where $O(k,\xi)$ are ordinary functions. Regarding these $O(k,\xi)$
as fourier components of a function $O(x,\xi)$, where $x^i$ are the
coordinates of a $2n$ dimensional space we then get the following
map between matrices and functions.
\begin{eqnarray}
& {\bf O} (\xi) \rightarrow  O(x,\xi);&~~~~[{\bf p}_i,{\bf O} (\xi) ]
=  i\partial_i O(x,\xi)\nonumber \\ 
&{\rm Tr}{\bf O} (\xi) =  {1
\over (2\pi)^{n}}[{\rm Pf~B}]\int d^{2n}x~~O(x,\xi)&
\label{eq:eleven}
\end{eqnarray} 
The product of two
matrices ${\bf O}_1 (\xi)$ and ${\bf O}_2 (\xi)$ is then mapped to a star
product
\begin{eqnarray}
{\bf O}_1(\xi) {\bf O}_2(\xi) 
& \rightarrow & O_1(x,\xi)*O_1(x,\xi), \nonumber \\
O_1(x,\xi)*O_2(x,\xi)
& \equiv & {\rm exp}~[{\theta^{ij}\over 2i}{\partial^2 \over \partial s^i
\partial t^j}]~O_1(x+s,\xi)O_2(x+t,\xi)~|_{s=t=0}
\label{eq:fourteen}
\end{eqnarray}
With these rules, one can easily verify that
\begin{eqnarray}
\bF_{\mu\nu} & \rightarrow &
\partial_\mu A_\nu - \partial_\nu A_\mu 
-i A_\mu *A_\nu + i A_\nu *A_\mu \equiv F_{\mu\nu}\nonumber \\ D_\mu
 \bX^i & \rightarrow & \theta^{ij}(\partial_\mu A_j -\partial_j A_\mu
 -i A_\mu * A_j + i A_j * A_\mu) \equiv \theta^{ij}F_{\mu j}\nonumber
 \\ D_\mu \bX^a & \rightarrow & \partial_\mu \phi^a - iA_\mu *
\phi^a  + i \phi^a *A_\mu \equiv D_\mu \phi^a\nonumber \\
~~[\bX^i,\bX^j] & \rightarrow &
i\theta^{ik}\theta^{jl}(F_{kl}-B_{kl})\nonumber \\ 
~~[\bX^i,\bX^a] &
\rightarrow & i\theta^{ij}(\partial_j \phi^a - iA_j \star \phi^a + i
\phi^a \star A_j) \equiv i\theta^{ij}D_j\phi^a
\label{eq:fifteen}
\end{eqnarray}
where we have defined
\begin{equation}
F_{ij} = \partial_i A_j-\partial_jA_i - 
iA_i\star A_j + iA_j\star A_i
\label{eq:sixteen}
\end{equation}
In the above equations the quantities appearing in the right hand side
are ordinary functions of $(x,\xi)$.

The action (\ref{eq:six}) becomes the
action of $U(1)$ noncommutative gauge theory in the $p+2n+1$ dimensions
spanned by $x,\xi$. The noncommutativity is entirely in the $2n$
directions. In addition to the gauge fields we also have $(9-p-2n)$
``adjoint'' scalars $\phi^a$. 
The gauge field appears in the combination
$F_{AB} - B_{AB}$
where $B_{AB}$ is an antisymmetric matrix whose $(ij)$ components are
$B_{ij}$ and the rest zero. This corresponds to a specific choice of
``description'' in the NCYM theory \cite{seibergwitten}. 
Furthermore the upper and
lower indices of various quantities some contracted with the ``open
string metric'' whose components in the nocommutative directions are
\begin{equation}
G^{ij} = - \theta^{ik}g_{kl}\theta^{lj}
\label{eq:seventeena}
\end{equation}
The componments of the open string metric in the commutative directions
are the same as the original metric $g_{ab}$. Finally the coupling
constant which appears in front is the open string coupling $G_s$ which
is related to the closed string coupling $g_s$ by
\begin{equation}
G_s = g_s ({{\rm det}(G-B)\over{\rm det}(g+B)})^{1\over 2}
= g_s ({{\rm det}~B\over{\rm det}~g})^{1\over 2}
\label{eq:seventeenb}
\end{equation}
It may be also easily verified that
\begin{equation}
{1\over G-B} = -\theta + {1\over g+B}
\label{eq:dilseven}
\end{equation}
(Recall that $\theta^{-1} = B$ as matrices.)

It is straightforward to extend the above construction to obtain a
nonabelian noncommutative theory. The classical solution which one
starts with is now
\begin{equation}
\bX^i (\xi) = \bx^i \otimes I_M
\label{eq:eighteen}
\end{equation}
where $I_M$ denotes the unit $M \times M$ matrix. Now the various
$\infty \times \infty$ matrices map on to $M \times M$ matrices which
are functions of $x$, in addition to $\xi$. 
With this understanding the formulae above can
be almost trivially extended. The star product would now include
matrix multiplication and the map for the trace becomes
\begin{equation}
{\rm Tr}{\bf O} (\xi) = {1
\over (2\pi)^{n}}[{\rm Pf~B}]\int d^{2n}x~~{\rm tr}~O(x,\xi)
\label{eq:nineteen}
\end{equation}
where ${\rm tr}$ now denotes trace over $M \times M$ matrices. Instead
of obtaining a $U(1)$ noncommutative theory one now obtains a
$U(m)$ noncommutative theory.

\section{Open Wilson lines}

Consider the following gauge invariant operator in the $Dp$-brane
theory \cite{iikk}
\begin{eqnarray}
W(C,k) & = & \int d^{p+1}\xi
~{\rm Lim}_{M \rightarrow \infty}~~
{\rm Tr}~[\prod_{n=1}^M~U_j]~e^{ik_\mu \xi^\mu}\cr
U_j & = & {\rm exp}~[i {\vec \bC} \cdot ({\vec \Delta d})_n]
\label{eq:twenty}
\end{eqnarray}
where $\Delta d_n$ denotes the $n$-th infinitesimal line element along
the contour $C$. The momentum components $k_\mu$ along the
commutative directions appear explicitly
in (\ref{eq:twenty}), while the components along the noncommutative directions
$k^i$ are given by
\begin{equation}
k_i = B_{ji}d^j
\label{eq:twoone}
\end{equation}
where $d^j$ are the components of the vector
${\vec d} = \sum_{n=1}^M {\vec{\Delta d}}$.
Applying the above dictionary it is easy to see that this operator
becomes the following expression in terms of the noncommutative gauge
field 
\begin{equation}
W (k,C) = \int d^{d+1}x~ {\rm tr}~[P_\star {\rm exp}[i\int_C d\lambda
 {d y^A(\lambda) \over d \lambda} A_A (x + y(\lambda))]~\star~e^{ik_B x^B}]
\label{eq:two}
\end{equation}
where we have used the indices $A,B = 1, \cdots (p+2n+1)$ to denote
all the directions collectively.  The trace in (\ref{eq:two}) is over the
nonabelian gauge group. $\lambda$ is a parameter that increases along
the path. In our conventions the path ordering is defined so that
points at later values of $\lambda$ occur successively to the
left. Equation (\ref{eq:twoone}) implies that the end points of the 
contour are separated by an
amount $\Delta x^A$ where
$\Delta x^A = k_B \theta^{BA}$. 
Clearly the separation is nonzero only along the noncommutative directions.

In a similar way consider the operator
\begin{equation}
\cO(k) = \int d^{p+1}\xi~e^{ik_\mu \xi^\mu}~{\rm Tr}~
[e^{ik_i\bX^i}~ \cO(\bX,\bA,\xi)]
\label{eq:twothree}
\end{equation}
where $\cO(\bX,\bA,\xi)$ is some operator in the $Dp$ brane gauge
theory which transforms according to the adjoint representation. In
terms of noncommutative gauge fields and star products, this becomes 
\begin{equation}
\tcO (k) = \int d^{d+1}x
~~{\rm tr}~~ \cO (x+ {k \cdot \theta}) \star P_\star~ {\rm exp}
[i\int_0^1 d\lambda~
k_A\theta^{AB} A_B (x + {k \cdot \theta~\lambda})]
 \star e^{ik\cdot x}.
\label{eq:four}
\end{equation}
The contour is now a {\it straight} path transverse to the
momentum  along the direction, $\eta^A  = k_B\theta^{BA}$.
$\cO (x + {k \cdot \theta})$ is a local operator constructed from the
fields which is inserted at the endpoint of the path, and
$(k \cdot \theta)^A \equiv k_B \theta^{BA}$.
This is the operator defined in \cite{ghi}. We will call the straight
Wilson line a Wilson tail.

The operator $\cO(\bX,\bA,\xi)$ can be a composite operator made of
field strengths,
$\bF_{\mu\nu}$, the covariant derivatives of the scalar fields $D_\mu\bX^I$
and $[X^I,X^J]$. That is
\begin{equation}
\cO (\bX,\bA,\xi)
= \prod_{\alpha=1}^n \cO_\alpha (\bX,\bA,\xi)
\label{eq:kofour}
\end{equation}
where each of the $\cO_\alpha$ denotes a $\bF_{\mu\nu}$, $D_\mu\bX^I$
or a $[X^I,X^J]$. A symmetrized trace (denoted by the symbol ``STr'') 
of the expression in(\ref{eq:twothree}) 
is then defined as follows. Imagine expanding the exponential in 
$e^{i k \cdot \bX}$. For some given term in the exponential we thus have a 
product of a number of $\bX$'s and $\cO_\alpha$'s. We 
symmetrize these various factors of $\bX$'s and $\cO_\alpha$'s 
and average.
The rules derived in the previous section
then lead to the following expression for the symmetrized trace
version of (\ref{eq:twothree}) :
\begin{equation}
{\hat \cO}(k)
= \int d^{d+1}x \int \prod_{\alpha=1}^n d\tau_\alpha~
~P_*~{\rm tr}~
[ \prod_{\alpha=1}^n 
O_\alpha(x^i + \theta^{ji}k_j\tau_\alpha) W_t(k,A,\phi,x)]  
\star e^{ik_ix^i}
\label{eq:wone}
\end{equation}
where $W_t$ denotes the Wilson tail
\begin{equation}
W_t (k,A,\phi,x) 
= {\rm exp} [i\int_0^1 d\lambda~
k_A\theta^{AB} A_B (x + {k \cdot \theta~\lambda})]
\label{eq:wthree}
\end{equation}
Thus the Wilson tail now has operators which are smeared over it. This
is the operator considered in \cite{liu}.

\section{The proposal}

Consider a large number of coincident $p$ branes with no $B$ field in
the presence of a weak supergravity background. Let us denote a
supergravity mode in momentum space by $\Phi (k_I,k_\mu)$ where
$k_\mu$ denotes the momentum along the brane and $k_I$ denotes the
momentum transverse to the brane. Let $X^I$ denote the transverse
coordinate and $\bA_\mu$ the gauge field on the brane. Then in the
brane theory, the transverse coordinates are represented by scalar
fields $\bX^I(\xi)$. Then the results of \cite{taylor,dbi} show that
the linearized coupling of some supergravity mode $\Phi (k_\mu,k_I)$
to this set of branes is of the form
\begin{equation}
\Phi (k_\mu,k_I) \int d^{p+1}\xi~e^{ik_\mu\xi^\mu}~{\rm STr}~
[e^{ik_I\bX^I}~\cO_\phi(\bX,\bA,\xi)] 
\label{eq:twofour}
\end{equation}
The exponential now contains the transverse matrices $\bX^I$ as well.

Our proposal is the following \cite{dastrivedi}. Once we know the
coupling (\ref{eq:twofour}) we can obtain the coupling of the same
supergravity mode to a $(p+2n)$ dimensional noncommutative brane by
simply expanding the matrices which appear around the relevant
classical solution as in (\ref{eq:nine}). Using the results of the
previous section, this coupling then becomes
\begin{equation}
S_{int}  = \Phi(k)\int {[d\xi dx] 
\over (2\pi)^n}
({\rm Pf}B)\int \prod_{\alpha=1}^n d\tau_\alpha e^{ik_\mu
\xi^\mu}~{\rm tr}P_\star [ \prod_{\alpha=1}^n O_\alpha(x^i +
\theta^{ji}k_j\tau_\alpha) W_t]
\star e^{ik_ix^i}
\label{eq:kofive}
\end{equation}
\begin{equation}
W_t(k,A,\phi) = 
{\rm exp}[i\int_0^1 d\lambda ~k_i\theta^{ij} A_j (x + \eta(\lambda))
+ i
\int_0^1 d\lambda ~k_a \phi^a(x + \eta (lambda))]
\label{eq:twofive}
\end{equation}
and $\eta^i(\lambda) = \theta^{ji}k_j\lambda$. This is the generalization
of the Wilson line with Higgs fields considered in \cite{dasrey}.

\section{Tests of the proposal : DBI approximation}

The proposal described above is quite general and makes no reference
to any approximation. However, the exact form of the
operator $\cO_\phi$ which appears in (\ref{eq:twofour}) is not known. These
operators are known in the low energy limit or in the DBI
approximation. We now perform a test of our proposal by considering
the dilaton coupling in 
the DBI approximation. For simplicity we consider a {\it single}
noncommutative {\it euclidean} $D(2n-1)$ brane 
(with $2n$ dimensional
worldvolume)with noncommutativity in all 
the directions. We will construct this from a large number $N$ of
$D(-1)$ branes. Following \cite{taylor,dbi} 
we will assume that the action in the
presence of a dilaton field $D(k)$ with momentum $k$ 
(with all backgrounds trivial)
is given by
\begin{equation}
S_{int} = {D(k)\over g_s} {\rm STr}~
e^{ik\cdot \bX}{\sqrt{{\rm det}
(\delta^I_J - i [\bX^I,\bX^K]g_{KJ})}}
\label{eq:diltwo}
\end{equation}
The notation is the same as in the previous sections. We then expand
around the classical solution as in (\ref{eq:seven}) and
(\ref{eq:eight}) to obtain a single noncommutative $(p+2n)$ brane.
To simplify things further we will take $k_a=0$ and also set $\phi^a =
0$. A nonzero $k_a$ or $\phi^a$ can be easily incorporated.

In the following we will be interested in terms upto $O(A^2)$ in the
noncommutative gauge fields. In the language of matrices we will be
interested in terms which contain at most two matrices. For such terms
there is no distinction between the symmetrized trace and ordinary
trace. We will therefore replace STr in (\ref{eq:diltwo}) with Tr.
Using the results of the previous section, this interaction is then written
in terms of noncommutative gauge fields $F_{ij}$ 
\begin{equation}
S_{int} = {D(k)\over G_s}
\int d^{2n}x~e^{ikx}~P_*[{\rm exp}~(i\int d\eta^i A_i (x + \eta(\lambda)))]
~{\sqrt{{\rm det}
(G + F - B)}}
\label{eq:dilfive}
\end{equation} 
where in (\ref{eq:dilfive}) the quantities $\theta,F,B,g$ are written as $(2n)
\times (2n)$ matrices and $I$ stands for the identity matrix, 
in a natural notation. In the following whenever these
quantities appear without indices they denote these matrices. We have used
(\ref{eq:seventeena}) and (\ref{eq:seventeenb}) to write the
expression
in terms of the open string
metric $G_{ij}$ and the open string coupling $G_s$.
Here the path used is given by (\ref{eq:five}) 
and all products are star products.

In terms of the ordinary gauge fields $f_{ij}$, the closed string metric
and the closed string coupling, the interaction may be
read off from the standard Dirac-Born-Infeld action
\begin{equation}
\tS_{int} = {D(k)\over g_s}
\int d^{2n}x~e^{ikx}~{\sqrt{{\rm det}
(g + f +B)}}
\label{eq:dilsix}
\end{equation} 
The strategy is now to express (\ref{eq:dilsix}) 
in terms of the noncommutative
gauge field $F_{ij}$ using the Seiberg-Witten map in a series involving
powers of the potential $A_i$ and compare the result with (\ref{eq:dilfive})
which is also expanded in a similar fashion. 

For zero momentum operators this is the comparison done in
\cite{seibergwitten}, 
where it is shown that
\begin{equation}
{1\over g_s}{\sqrt{{\rm det}
(g + f +B)}} = {1\over G_s}{\sqrt{{\rm det}
(G -B + F)}} + O(\partial F) + {\rm total~derivatives}
\label{eq:dilsixa}
\end{equation}
which shows the equivalence of the two actions in the presence of 
constant backgrounds. The crucial aspect of our comparison is the presence
of these total derivative terms in (\ref{eq:dilsixa}) which cannot be ignored
if $k \neq 0$. We will find that these total derivative terms are
in precise agreement with similar terms coming from the expansion of the
Wilson tail in (\ref{eq:dilfive}) upto $O(A^2)$.

Since we are using the DBI action, the field strengths should be
really treated as constant.  In carrying out the comparison, however,
some caution must be exercised.  Since the Seiberg-Witten map contains
gauge potentials as well as field strengths a term containing a
derivative of a field strength multiplied by a gauge potential without
a derivative on it, cannot be set automatically to zero, as emphasised
in \cite{seibergwitten}.

The details of this comparsion is given in \cite{dastrivedi}. The fact
that operators with Wilson tails can be written in terms of
generalized star products \cite{star} becomes very useful in the
calculations. In particular the expansion of the expression
(\ref{eq:wone}) to $O(A^2)$ is
\begin{eqnarray} 
& {\hat \cO }(k) =
\int d^{p+1}\xi {d^{2n}x \over
(2\pi)^n} e^{ik_\mu\xi^\mu}~({\rm PfB}){\rm tr}~[\cO (x,\xi)
& + \theta^{ij}\partial_j(\cO \star^\prime A_i) \nonumber \\
& & +  {{1\over 2}} \theta^{ij}
\theta^{kl}\partial_j\partial_l[\cO A_i A_k]_{*3}]\star
e^{ik_ix^i}
\label{eq:pleight}
\end{eqnarray}
The various generalized star products are defined in \cite{star}.

Consider first the comparison to 
$O(A)$. To this order the 
Seiberg-Witten map simply reduces to
$f_{ij} = \partial_i A_j - \partial_j A_i + O(A^2)$.
Thus it is sufficient to expand the determinant in (\ref{eq:dilsix}) 
to linear order in
$f$. One obtains
\begin{equation}
\tS^{(1)}_{int} = { D(k) \over g_s}\sqrt{{\rm det}
(g + B)}~\int d^{2n}x ~e^{ikx}~
[1 + {1\over 2} ({1\over (g+B)})^{ij}(\partial_j A_i 
- \partial_i A_j) + O(A^2)]
\label{eq:dilnine}
\end{equation}
Using (\ref{eq:seventeenb}) and (\ref{eq:dilseven}) this may be written as
\begin{equation}
\tS^{(1)}_{int} = { D(k) \over G_s}
\sqrt{{\rm det}(G - B)}~\int d^{2n}x ~e^{ikx}~[1 +
{1\over 2} ({1\over (G-B)} + \theta )^{ij}(\partial_j A_i 
- \partial_i A_j)+ O(A^2)]
\label{eq:dilten}
\end{equation}
The products in all expressions which involve $A$ and $F$ rather than 
$a$ and $f$ are star products.
We have to compare this with the expansion of the expression
(\ref{eq:dilfive}) to $O(A)$. 
To do this we can use (\ref{eq:pleight}) with the function
$\cO$ being replaced by the quantity ${\sqrt{{\rm det}
(G + F - B)}}$. To linear order in $A$ we have
\begin{equation}
{\sqrt{{\rm det}
(G + F - B)}}= {\sqrt{{\rm det}(G-B)}}[1 + {1\over 2}
({1\over G-B})^{ij} (\partial_i A_j 
- \partial_j A_i) + O(A^2)]
\label{eq:dileleven}
\end{equation}
The various products appearing on the left hand side of
the above equation are star products. However to this order
these collapse to ordinary products since $G,B$ etc. are
constants.
Also to this order one has
\begin{equation}
\theta^{ij}\partial_j ({\sqrt{{\rm det}
(G + F - B)}} \sp A_i) = 
{{1\over 2}} {\sqrt{{\rm det}
(G - B)}} \theta^{ij}(\partial_j A_i - \partial_i A_j)
\label{eq:diltwelve}
\end{equation}
Using (\ref{eq:dileleven}) and (\ref{eq:diltwelve}) it is easy to see
that
$S_{int}$ agrees with (\ref{eq:dilten}) to $O(A)$.
Note that the term proportional to $\theta$ in (\ref{eq:dilten}) came
because of the relation (\ref{eq:dilseven}) while the corresponding
term on the noncommutative side came from the ``Wilson tail'' involved
in the gauge invariant operator. 
To this order one is sensitive only to the linear term
of the Seiberg Witten map.  However the agreeement
of the two derivations of the interaction term is still nontrivial
and the importance of the open Wilson line is evident. 

The nontriviality of the Seiberg-Witten map enters at $O(A^2)$. Using
the various properties of generalized star products in \cite{star} and
folowing the same strategy as above we have checked that the operator
we propose is indeed identical to the standard operator in terms of
ordinary gauge fields. See \cite{dastrivedi} for details.
While we have not carried out the calculation explicitly to $O(A^3)$ we are
fairly confident that the agreement will persist. At this order the
nontriviality of the symmetrized trace will be important.

\section{Other developments}

Another test of our proposal can be performed in the Seiberg-Witten
low energy limit, where the results can be compared with explicit
string worldsheet calculations.  This has been done in
\cite{worldsheet} and the results are in agreement for superstrings,
though not for bosonic strings. For bosonic theory, however, we do not
expect the above procedure to work since the nontriviality of the
matrix model measure will swamp whatever classical solution we start
with.

A recent interesting development is the determination of the operators
which couple to RR fields \cite{rrcoupling}, generalizing earlier
results on couplings to constant RR fields \cite{nemmuk1}.
One of the outcomes is an exact Seiberg-Witten
map. Another recent development has been a proposal that closed string
fields can be reconstructed out of open Wilson lines of arbitrary
shapes by a suitable harmonic expansion \cite{dkit2}.

\section{Holography}

So far we have considered the linearized couplings of noncommutative branes
to supergravity modes. Another context in which gauge invariant
operators should arise is in holography. Supergravity duals of
noncommutative gauge theories are known \cite{hi,mr}. Supergravity
modes in these backgrounds should be dual to momentum space gauge
invariant operators in the gauge theory \cite{dghosh}. 
Generally, these operators are not identical to the operators we have
considered, though for some cases they can be obtained by linearizing
e.g. the DBI action around the background geometry
\cite{dtr}. Moreover, as argued in \cite{taylor}, it is possible to
obtain the correlation functions of these operators from those of the
operators which couple to linearized backgrounds by solving the
scattering problem in the full geometry.

The identification of the holographically dual operators in this
context is an open problem. However there is one observation which may
be useful. The asymptotic geometry for the $p+2n+1$ dimensional
non-commutative theory is {\it identical} to that for the $p+1$
dimensional ordinary theory at a particular point in the Coulomb
branch where the $p$-branes are spread out uniformly along the $2n$
directions.  This is in fact the dual manifestation of the
relationship between commutative and noncommutative Yang-Mills
theories discussed above \cite{roylu}.
This connection may be possibly used to 
tackle the problem of mode mixing in such supergravity backgrounds.
We do not have definitive results about this at present.

These holographically dual
operators should also involve Wilson tails. A good indication is the
fact that the linear relationship between the size of these operators
and the momentum is something which is also visible in the dual
supergravity theory \cite{dasrey}. 
In the AdS/CFT correspondence, the size of the
hologram of an object in the bulk {\em decreases} monotonically as the
object moves closer to the boundary. In the supergravity duals of
noncommutative gauge theories, something interesting happens. In these
backgrounds, the region deep into the bulk is $AdS$ space-time and the
above IR/UV relationship holds. However near the boundary, the
relationship is {\em opposite} \cite{dasrey}. In this region the size
of the hologram {\em increases} as the object moves closer to the
boundary. A similar relationship was found in the full $D3$ brane
geometry in \cite{daniel}. The latter is known to be a special case of
the former \cite{dghosh}. 
The relationship between the hologram size and the momentum
scale of the NCGT is in fact almost the same as that between the size
of the Wilson tails and the momentum : they differ by a factor of the
square root of the 't Hooft coupling. The latter factor, is however,
always present in the relationship between the noncommutativity scale
of the gauge theory and the scale observed in the supergravity dual 
and is presumably a strong coupling effect.

While this is a good indication, the precise form of the dual
operators remains to be discovered. A knowledge of these operators
should throw valuable light on holography in backgrounds which are not
asymptotically $AdS$.

\bibliographystyle{amsalpha}

\end{document}